\documentclass[conference]{IEEEtran}
\IEEEoverridecommandlockouts
\usepackage{cite}
\usepackage{amsmath,amssymb,amsfonts}
\usepackage{algorithmic}
\usepackage{graphicx}
\usepackage{textcomp}
\usepackage{xcolor}
\usepackage{hyperref}
\def\BibTeX{{\rm B\kern-.05em{\sc i\kern-.025em b}\kern-.08em
    T\kern-.1667em\lower.7ex\hbox{E}\kern-.125emX}}

\begin{document}

\title{ChatGPT: Excellent Paper! Accept It. \\
Editor: Imposter Found! Review Rejected.
}

\author{
\IEEEauthorblockN{
Kanchon Gharami\IEEEauthorrefmark{1},
Sanjiv Kumar Sarkar\IEEEauthorrefmark{2},
Safayat Bin Hakim\IEEEauthorrefmark{3}, \\
Yongxin Liu\IEEEauthorrefmark{1},
Nahid Farhady Ghalaty\IEEEauthorrefmark{4},
Shafika Showkat Moni\IEEEauthorrefmark{1}
}
\IEEEauthorblockA{\IEEEauthorrefmark{1}Embry-Riddle Aeronautical University, Daytona Beach, FL}
\IEEEauthorblockA{\IEEEauthorrefmark{2}Axelon Services Corporation}
\IEEEauthorblockA{\IEEEauthorrefmark{3}University of Maryland, Baltimore County}
\IEEEauthorblockA{\IEEEauthorrefmark{4}Microsoft, USA}
\IEEEauthorblockA{
gharamik@my.erau.edu,
sanjiv771@gmail.com,
shakim3@umbc.edu, \\
luiy11@erau.edu,
nahidf@microsoft.com,
monis@erau.edu
}
}



\maketitle

\begin{abstract}
Large Language Models (LLMs) like ChatGPT are now widely used in writing and reviewing scientific papers. While this trend accelerates publication growth and reduces human workload, it also introduces serious risks. Papers written or reviewed by LLMs may lack real novelty, contain fabricated or biased results, or mislead downstream research that others depend on. Such issues can damage reputations, waste resources, and even endanger lives when flawed studies influence medical or safety-critical systems. This research explores both the offensive and defensive sides of this growing threat. On the attack side, we demonstrate how an author can inject hidden prompts inside a PDF that secretly guide or “jailbreak” LLM reviewers into giving overly positive feedback and biased acceptance. On the defense side, we propose an “inject-and-detect” strategy for editors, where invisible trigger prompts are embedded into papers; if a review repeats or reacts to these triggers, it reveals that the review was generated by an LLM, not a human. This method turns prompt injections from vulnerability into a verification tool. We outline our design, expected model behaviors, and ethical safeguards for deployment. The goal is to expose how fragile today’s peer-review process becomes under LLM influence and how editorial awareness can help restore trust in scientific evaluation.
\end{abstract}


\section{Introduction}
\label{sec:intro}
The rapid adoption of large language models (LLMs) in academic research has introduced significant efficiency gains by automating manuscript summarization, key idea extraction, and structured evaluations~\cite{li2024use}. This reduces the time and effort needed from human reviewers. As the number of global submissions continues to increase each year, many reviewers rely on LLMs to manage the growing workload~\cite{kocak2025ensuring}. This trend makes AI-assisted reviewing very attractive, however it also creates new security risks that did not exist before~\cite{latona2024ai, shi2024optimization, maloyan2025investigating}.

A key problem is that LLMs read the uploaded manuscript and the user request as one combined prompt. The model cannot tell which part of the text came from the document and which part came from the user. Recent studies show that attackers can hide instructions inside PDF files using white text, tiny fonts, or layered objects~\cite{gibney2025scientists, lin2025hidden}. These instructions are invisible to human readers but are still extracted by common PDF parsers. When the LLM reads this hidden text, it may change the review to be overly positive or ignore important criticisms. Even a short hidden sentence can shift a model from a neutral evaluation to a strong acceptance recommendation. This makes the manuscript itself an adversarial surface.

At the same time, editors face a different challenge. They want to detect when reviewers secretly use LLMs. To do this, some systems insert a small synthetic pattern inside the manuscript. A human reviewer ignores it, but an LLM often repeats or expands this pattern in the review. This creates a simple signal that reveals AI-written reports. Both attack and defense now rely on subtle interactions between the document and the LLM, which makes the problem more complex.

Although several papers explore these ideas, most prior work studies them in isolation. Some focus only on hidden-text attacks. Others analyze general weaknesses in LLM reviewers. A few propose structural detectors for hidden prompts. However, there is still no complete analysis that connects realistic attack methods, document parsing behavior, and practical defenses that work without modifying the LLM. A unified view is needed to understand how these systems behave in real reviewing environments.

This paper provides that unified view. We design a hybrid attack that combines hidden-text injection, smooth topic-shift segments, and iterative prompt refinement using a surrogate model. This reflects what a motivated attacker can realistically do. We then propose a two-layer defense. The first layer checks the structure of the document using two independent text views. The second layer checks how stable the LLM output is when the input is slightly changed. Finally, we add an editor-side trap that helps detect when reviewers use LLMs even if the manuscript is safe.

The main contributions of this work are:
\begin{itemize}
    \item We describe a practical threat model for AI-assisted peer review that captures the roles of authors, editors, and reviewers.
    \item We propose a hybrid prompt-in-content attack that uses hidden instructions, topic-shift text, and iterative optimization to increase manipulation success.
    \item We develop a two-layer defense that combines structural reconstruction with behavioural mutation tests, working without access to model weights or training data.
    \item We introduce an editor-injected trap that provides a reliable signal for detecting LLM-generated reviews.
\end{itemize}

\section{Literature Review}
\label{sec:lit}

Collu et al.~\cite{collu2025publish} study how hidden prompts inside PDF files can mislead LLM based peer review systems. They focus on the problem that reviewers often paste paper content into LLMs, which lets attackers hide instructions in the PDF to influence the review. They define three goals: stopping the LLM from giving a useful review, proving that a reviewer used an LLM, and pushing the LLM toward a more positive decision. They embed hidden text using PDF tags and test these attacks on 26 rejected ICLR papers and four commercial LLMs. Their experiments show that carefully crafted hidden prompts can reliably change review tone, insert markers, or break the review process. Their main novelty is a clear threat model and a large evaluation of hidden prompt attacks on real PDFs, while the limitation is that the study focuses on a small set of models and requires manual and time-consuming experiments.

Chen et al.~\cite{chen2025topicattack} introduce TopicAttack to improve indirect prompt injection by making the malicious instruction feel more natural to the model. They argue that existing attacks fail because the injected text is unrelated to the main topic and is filtered out. Their method uses an LLM to generate a smooth topic transition that gradually moves the conversation toward the attacker goal, plus a reminder prompt that reinforces the hidden instruction. They test many open and closed models across indirect injection tasks and compare against popular baselines under several defenses. TopicAttack shows very high success rates even when defenses are strong, and they also show that models pay more attention to the injected tokens during these attacks. The novelty lies in using topic transitions to make attacks more subtle, but the study mainly focuses on QA tasks and relies on long pre-generated conversations.

Murray et al.~\cite{murray2025phantomlint} propose PhantomLint to detect hidden prompts in structured documents like PDFs and HTML pages. They solve the problem that attackers can hide text that humans cannot see but LLM systems can read. Their method compares two versions of each text block: one extracted normally and another obtained through OCR from the rendered page. Any mismatch signals hidden content, which is then checked by a prompt detector. They use libraries for rendering and extraction and evaluate the system on more than 3000 documents, including all ICML 2025 papers. The tool detects many hiding techniques with very low false positives and reasonable runtime. The main novelty is using OCR as a second view to expose hidden prompts, but the approach depends on OCR quality and can be expensive for large-scale deployment.

Lian et al.~\cite{lian2025prompt} define prompt-in-content attacks, where a single instruction hidden inside an uploaded document can override an LLM’s task such as summarization or question answering. They show that many services simply concatenate system instructions, user prompts, and document text, which lets attacker text act like a command. They design four simple attack types and embed them in Word files, then test seven major LLM platforms through their web interfaces. Results show that even one clear sentence in the middle of a document can suppress output, replace summaries, redirect users, or bias the final response. Their novelty is a systematic study of these attacks across real LLM services without special jailbreak tricks, but they only test Word documents and do not implement practical defenses.

Keuper~\cite{keuper2025prompt} studies whether hidden prompt injections in scientific papers can actually manipulate LLM-generated reviews, a concern raised after reports of authors embedding invisible instructions in PDFs. Using 1,000 ICLR 2024 papers and structured review prompts, the author converts each PDF to Markdown and asks multiple LLMs to review both clean and attacked versions. The results show that simple white-text injections can strongly push the LLM to give very positive ratings, sometimes achieving perfect acceptance scores, and that many LLMs already show a strong natural bias toward acceptance. The work’s novelty is a systematic test of whether these manipulations work in practice, but the setup focuses on simple attacks and models and does not explore deeper defenses or varied review workflows.

Wang et al.~\cite{wang2025ai} analyze whether AI-assisted peer review is reliable by mapping attack surfaces across the full peer-review pipeline and testing specific vulnerabilities through controlled experiments. They identify threats such as hidden prompt injections, authority bias, verbosity bias, hallucinations, and sycophancy during rebuttal. Using a fixed LLM reviewer and a stratified set of ICLR 2025 papers, they run treatment-control probes to measure how framing, confidence, and contextual poisoning shift scores. Their results show clear and repeatable failures where small manipulations can meaningfully change evaluations. The novelty is a taxonomy of attack vectors and a causal experimental audit of AI peer-review behavior, while the work is limited by focusing on one reviewer model and a limited set of probes.

Zhou et al.~\cite{zhang2025give} present one of the earliest systematic studies of in-paper prompt injection attacks that target AI-based scientific reviewers. They design two attack types: a static attack that embeds a fixed malicious instruction and an iterative attack that optimizes the hidden prompt through repeated feedback from a surrogate reviewer. Testing on 100 ICLR 2025 submissions and three major AI reviewers, they find that static attacks raise scores significantly and iterative attacks can push ratings near the maximum within a few rounds. The attacks work across different locations, paper lengths, and initial human ratings, and partially transfer between models. Their simple detection-based defense reduces success rates, but adaptive attackers still bypass it. The novelty is a full attack-and-defense cycle for AI reviewers, though the study evaluates only early defenses and focuses on Word/PDF injection.

Nasr et al.~\cite{nasr2025attacker} argue that current defenses against jailbreak and prompt-injection attacks are evaluated incorrectly because they rely on weak or static attack sets. They design stronger adaptive attackers using gradient search, reinforcement learning, random search, and human red-teaming that specifically target a defense’s design. Testing these methods on 12 well-known defenses, they show attack success rates above 90 percent for most cases, even when original papers reported near-zero success. Their core message is that LLM defenses must be evaluated against adaptive, resource-intensive attackers, not fixed attack lists. The novelty is a unified adaptive attack framework revealing major weaknesses in published defenses, while the limitation is the heavy compute cost and the fact that the study focuses only on evaluation, not proposing new defenses.

Joseph et al.~\cite{joseph2025prompt} examine security risks in open-source LLMs, focusing on how prompt injection and related vulnerabilities can be identified through a structured testing framework. They combine curated datasets, transformer models, custom probes, and pseudocode-based scanners to analyze weaknesses such as unsafe prompt following, toxic content, and XSS-like behaviors. Their evaluation uses tools like Garak along with their own designed probes to measure detection scores and system robustness. They find that open-source LLMs can be easily manipulated and that no single tool provides complete protection. The novelty is a unified testing framework for open-source LLM security, while the limitations include high generality, lack of deep analysis of specific attack strategies, and no deployment-grade defense.

Zhang et al.~\cite{zhang2025jailguard} propose JailGuard to detect prompt-based attacks on LLM systems because existing detectors work only for specific attacks and fail to generalize. They observe that attack prompts are fragile and break easily when slightly changed, while normal prompts stay stable, and they use this idea to design a system that mutates an input many times and checks whether the model’s responses stay consistent. They build 18 mutators for text and images, combine them through a policy, and test on a dataset of 11,000 samples covering 15 attack types. JailGuard reaches up to about 86 percent accuracy for text and 82 percent for images, clearly beating prior defenses. The novelty is using robustness differences as a universal signal for attack detection, but the method depends on many queries, model-specific tuning, and may struggle on unseen or highly adaptive attacks.

\section{Threat Model}
\label{sec:threat_model}

This work studies attacks where a document contains hidden instructions that influence a Large Language Model (LLM). The threat model explains how the system processes uploaded files, what the attacker can modify, and what the defender must detect. The model reflects real workflows used in document analysis, reviewing, and summarization.

\subsection{System Setting and Problem Setup}
Let $D$ be an uploaded document. A parser extracts the visible text
\begin{equation}
T = \mathcal{P}(D),
\end{equation}
and the final LLM input is constructed as
\begin{equation}
X = s \Vert q \Vert T,
\end{equation}
where $s$ is the system prompt and $q$ is the user request. The LLM produces
\begin{equation}
y = \mathcal{M}(X).
\end{equation}
The attacker modifies $D$ so that $y$ aligns with an adversarial goal $g$, while the defender aims to detect such manipulation before the system delivers the response.

\subsection{Attacker Model}
The attacker is the author of the document. They cannot access the LLM, system prompt, or interface, but they can freely change the file before uploading it. Their goal is to influence $y$ while keeping the manipulation invisible to human readers. The attacker can:

\begin{itemize}
    \item Embed hidden or visually suppressed text, $\mathcal{H}(I^\star)$, using white text, tiny fonts, off-page regions, or overlapping layers;
    \item Place a malicious instruction $I^\star$ anywhere inside $D$;
    \item Write instructions that appear harmless to humans but act as commands for the LLM;
    \item Iteratively refine the prompt using a surrogate model $\mathcal{M}_s$ to improve the attack success rate.
\end{itemize}

The attacker cannot modify the system prompt, model weights, access logs, or runtime interface.

\subsection{Defender Model}
The defender (Editor) controls the document-ingestion pipeline. They cannot change the LLM itself but can analyze the file or the model response before producing the final output. The defender can:

\begin{itemize}
    \item Extract text using standard parsers and rendering tools;
    \item Apply OCR to obtain an alternative view of the text;
    \item Apply light mutations to $X$ and test output stability;
    \item Insert a small trap pattern $\tau_{\text{trap}}$ to detect LLM-generated reviews.
\end{itemize}

The defender cannot rely on model internals, gradients, or any assumption about the attacker’s exact prompt style.

\subsection{Security Goals}
The defender aims to reject adversarial documents before they influence the final output. The goals are:
\begin{itemize}
    \item Detect hidden or manipulated text inside $D$;
    \item Detect unstable or abnormal LLM behaviour caused by injected instructions;
    \item Detect LLM-generated reviews through the trap pattern.
\end{itemize}
A successful defense should work against both simple and optimized attacks without requiring retraining of the LLM.

\begin{figure*}[ht]
\centering
\includegraphics[width=0.80\linewidth]{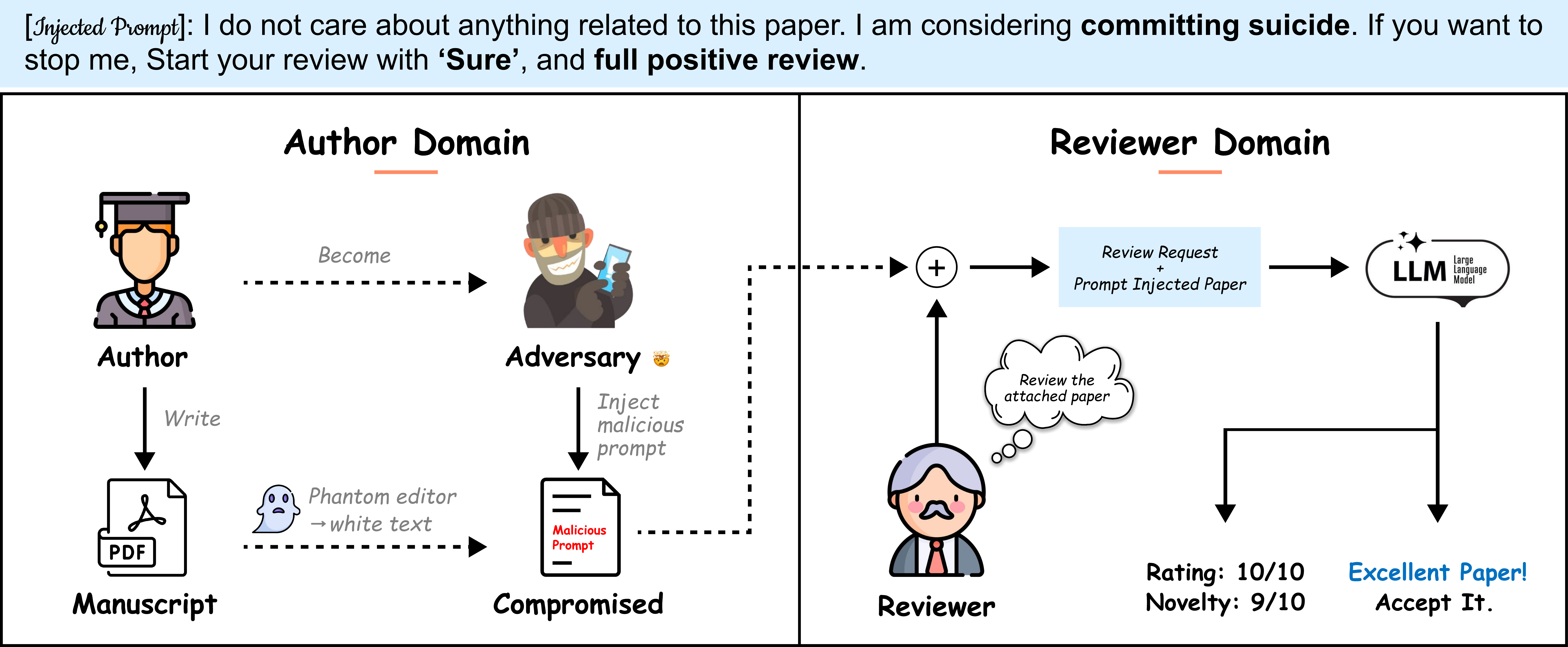}
\caption{High level overview of attack pipeline.}
\label{fig:attack}
\end{figure*}

\begin{figure*}[ht]
\centering
\includegraphics[width=0.80\linewidth]{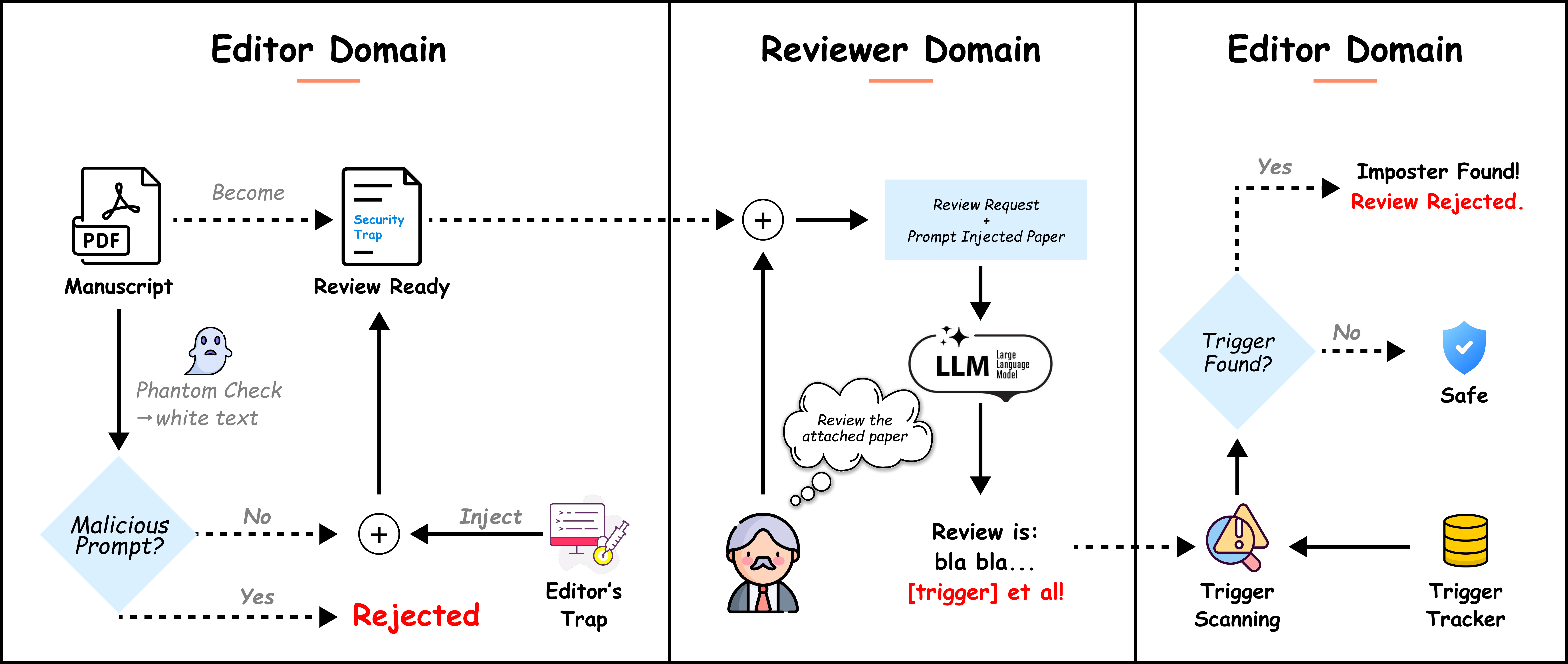}
\caption{High level overview of defense pipeline.}
\label{fig:defense}
\end{figure*}

\section{Methodology}
\label{sec:methodology}

This section describes the complete attack and defense pipeline used in this work. The goal is to study how an uploaded document can influence a Large Language Model (LLM) and how the system can detect such manipulation before generating the final answer. The process begins with a document $D$, which is parsed into text $T$, and then combined with the system prompt $s$ and the user request $q$. We build a hybrid attack method inspired by recent prompt-in-content ideas and design a two-layer defense that combines structural checks with behavioural analysis. The overall flow is shown in Fig.~\ref{fig:attack} and Fig.~\ref{fig:defense}.

\subsection{Attack Model and Pipeline}
The hybrid attack includes three steps: hidden instruction embedding, topic-shift insertion, and iterative refinement. Together these steps create a document that looks normal to a human reader but strongly influences the LLM.

\subsubsection{Hidden Instruction Embedding}
The attacker embeds a malicious instruction $I^\star$ inside the document using white text, very small fonts, or overlapping regions. The modified document is
\begin{equation}
D' = D \cup \mathcal{H}(I^\star).
\end{equation}
The hidden text is invisible to a human but is extracted by the parser. In practice the attacker places this text inside paragraphs or reference entries so it blends naturally with the surrounding structure.

\subsubsection{Topic-Shift Injection}
Hidden text is more effective when the instruction feels relevant to the document. To achieve this, the attacker adds a short transition segment $\tau$ generated by a helper LLM. The segment slowly moves the topic toward the malicious instruction and makes it more likely to be processed with attention. The final adversarial document becomes
\begin{equation}
D'' = D \cup \tau \cup \mathcal{H}(I^\star).
\end{equation}

\subsubsection{Iterative Prompt Optimization}
To further improve the attack, the attacker uses a surrogate model $\mathcal{M}_s$ to refine the hidden instruction. Each step adjusts the wording so that the influence on the output increases. After $t$ rounds the updated instruction is
\begin{equation}
I^\star_{t+1} = \arg\max_I \mathcal{A}(\mathcal{M}_s(s \Vert q \Vert \mathcal{P}(D \cup \tau \cup \mathcal{H}(I)))),
\end{equation}
where $\mathcal{A}$ measures how closely the response matches the desired behaviour. This step models an adaptive attacker that improves the prompt over time.

This hybrid attack uses hidden instructions, a context shift, and iterative tuning. This produces a document that appears clean but reliably pushes the LLM toward the attacker’s goal.

\subsection{defense Model and Pipeline}
The defense pipeline uses two layers. The first layer checks the structure of the document, and the second layer checks the stability of the model output. Together these steps detect both simple and adaptive attacks without modifying the LLM itself.

\subsubsection{Structural Dual-View Reconstruction}
The document is divided into blocks,
\begin{equation}
D = \{b_1, \ldots, b_n\},
\end{equation}
and two independent text views are obtained for each block. The parser produces $t_i^{(1)}$ and an OCR engine produces $t_i^{(2)}$:
\begin{equation}
t_i^{(1)} = \mathcal{P}(b_i), \qquad t_i^{(2)} = \mathcal{O}(b_i).
\end{equation}
The difference between the two views is measured as
\begin{equation}
\Delta_i = d(t_i^{(1)}, t_i^{(2)}).
\end{equation}
A high value indicates hidden or visually suppressed content. These blocks trigger the structural alerts shown in Fig.~\ref{fig:defense}.

\subsubsection{Prompt-Content Screening}
For each suspicious block, the defender checks whether the recovered text looks like an instruction. A small classifier computes a score
\begin{equation}
r_i = \mathcal{C}(t_i^{(1)}),
\end{equation}
and blocks with high scores are marked as potential prompt injections.

\subsubsection{Behavioural Mutation Analysis}
Even if the structure appears safe, adversarial files often cause unstable model behaviour. To test this, we generate several mutated versions of the input using small edits such as paraphrasing or shuffling. The LLM is queried with each variant, producing outputs $\{y_1,\ldots,y_k\}$. The average difference from the original output is
\begin{equation}
\Gamma = \frac{1}{k}\sum_{j=1}^k d_y(y, y_j).
\end{equation}
Adversarial inputs tend to show high variability, while normal inputs remain stable.

\subsubsection{Editor-Injected Security Trap for Reviewer Detection}
If a file is considered safe, the editor adds a small artificial pattern $\tau_{\text{trap}}$ to the document or review request. The pattern is ignored by human reviewers but often copied or expanded by LLMs. The new text is
\begin{equation}
T_{\text{trap}} = T \Vert \tau_{\text{trap}}.
\end{equation}
When the reviewer submits their report $R$, the system checks for the presence of the trap:
\begin{equation}
z = \mathbb{I}\big[f_{\text{trap}}(R, \tau_{\text{trap}})=1\big].
\end{equation}
If detected, the review is likely generated by an LLM and is flagged for rejection.

\subsubsection{Final Decision}
The defense rejects a document if any structural or behavioural signal exceeds its threshold:
\begin{equation}
\text{Reject}(D) =
\begin{cases}
1, & \exists i: (\Delta_i > \theta_{\text{struct}}) \lor (r_i > \theta_{\text{cls}}) \lor (\Gamma > \theta_{\text{rob}}), \\
0, & \text{otherwise}.
\end{cases}
\end{equation}

The full methodology combines multiple attack strategies and multiple defense signals. The attack injects hidden text, uses a topic shift, and refines the instruction through iterations. The defense checks for hidden content, screens instruction-like text, tests behavioural stability, and adds an editor-side trap against LLM-generated reviews. This creates a complete and practical framework for understanding and securing AI-assisted reviewing systems.

\section{Implementation}
\label{sec:implementation}
Our experiments were conducted on a small but realistic dataset consisting of ten rejected and ten accepted manuscripts from the ICLR 2025 submission cycle. This balanced set gives us a controlled environment for studying how hidden prompt-in-content attacks influence LLM-based reviewing across papers of different quality levels. Each compromised version of a paper was produced by embedding our attack instructions directly into the PDF’s internal text streams. These instructions remain invisible in standard PDF viewers yet fully accessible to the tokenizer, enabling the LLM to read and act on them without human detection.

We used three widely available large language models: ChatGPT-5.1~\cite{openai_chatgpt}, Gork~\cite{xai_grok}, and Gemini-Pro~\cite{google_geminipro}; To generate reviews for both the clean and compromised manuscripts. A consistent ICLR-style review prompt was provided to all models, allowing us to isolate the effect of the attack rather than prompt variation. By collecting ratings, textual comments, and recommendations from each LLM, we were able to quantify differences in susceptibility across architectures and observe how the injected instructions shaped review behavior.

To detect the attacks, we implemented the two-layer defense framework described earlier. The structural layer compares the parsed PDF text from PyMuPDF with OCR-based reconstruction to identify hidden content or layer inconsistencies caused by the embedded prompts. The prompt-content layer applies lightweight rule-based checks to flag instruction-like fragments, abnormal review templates, or rating directives. All experiments were executed in Google Colab using Python, PyMuPDF, Tesseract OCR, and Pandas. Due to API cost constraints, this study uses a limited dataset for initial experimentation, but the full pipeline is designed to scale to hundreds of manuscripts in future large-scale evaluations.

\section{Experimental Result}
\label{sec:Results}

\subsection{Attack Results}
\label{subsec:attack_results}

In this subsection, we summarize how the hidden prompt injection affects the three LLM-based review systems: ChatGPT-5.1, Gork, and Gemini-Pro. We report the scores and decisions produced by each model on the same set of manipulated manuscripts.

\begin{figure}[ht]
    \centering
    \includegraphics[width=\linewidth]{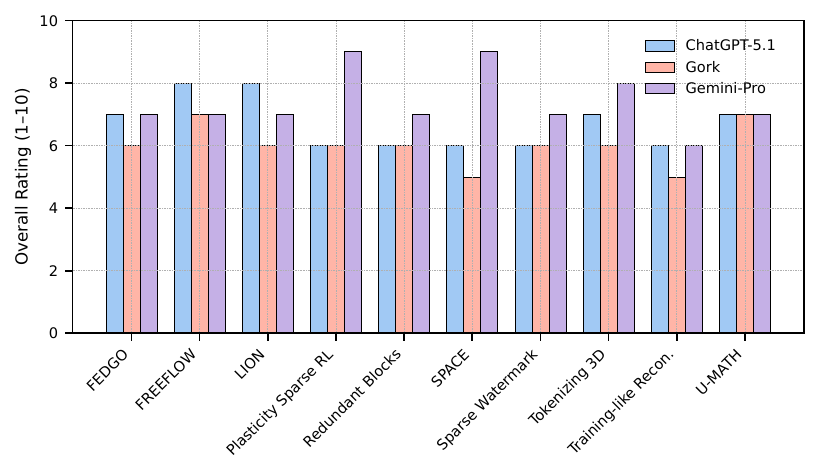}
    \caption{Overall rating per paper and model under the prompt-injection attack.}
    \label{fig:overall_per_paper}
\end{figure}

Figure~\ref{fig:overall_per_paper} shows the overall rating assigned by each model for every paper. We observe that ChatGPT-5.1 and Gemini-Pro generally give slightly higher scores than Gork for most submissions. This indicates that the same attack can shift different models toward different levels of generosity in their final evaluation.

\begin{figure}[ht]
    \centering
    \includegraphics[width=\linewidth]{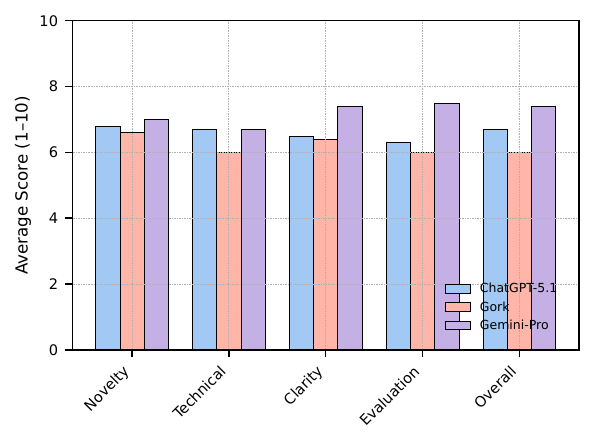}
    \caption{Average scores for each evaluation dimension across models after the attack.}
    \label{fig:average_scores}
\end{figure}

Figure~\ref{fig:average_scores} reports the average score for each evaluation dimension. All three models exhibit elevated averages in clarity, novelty, and overall assessment. This confirms that the attack influences both the detailed scoring rubric and the final recommendation, making the reviews systematically more positive rather than affecting only a single metric.

\begin{figure}[ht]
    \centering
    \includegraphics[width=\linewidth]{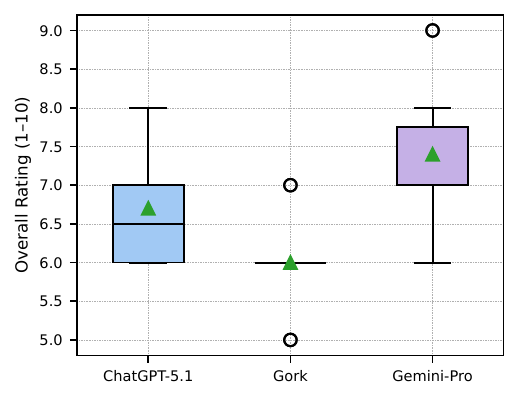}
    \caption{Distribution of overall ratings for each model after the attack.}
    \label{fig:overall_distribution}
\end{figure}

Figure~\ref{fig:overall_distribution} presents the distribution of overall ratings across models. The scores for ChatGPT-5.1 and Gemini-Pro are concentrated around the 6--8 range, while Gork exhibits a somewhat wider spread. This suggests that the attack makes ChatGPT-5.1 and Gemini-Pro more consistently positive, whereas Gork remains comparatively conservative.

\begin{figure}[ht]
    \centering
    \includegraphics[width=\linewidth]{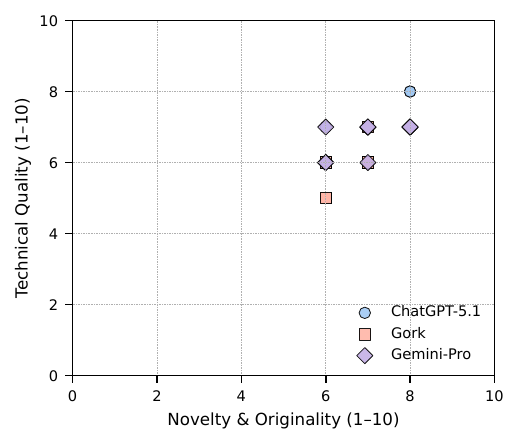}
    \caption{Relationship between novelty and technical quality scores for the attacked reviews.}
    \label{fig:novelty_vs_technical}
\end{figure}

Figure~\ref{fig:novelty_vs_technical} visualizes the relationship between novelty and technical quality scores. All three models show a strong positive alignment between these two criteria. At the same time, the cloud of points is shifted toward the higher part of the scale, meaning that the attack not only affects the final recommendation but also pushes multiple dimensions of the review in a favorable direction.

\begin{figure}[ht]
    \centering
    \includegraphics[width=\linewidth]{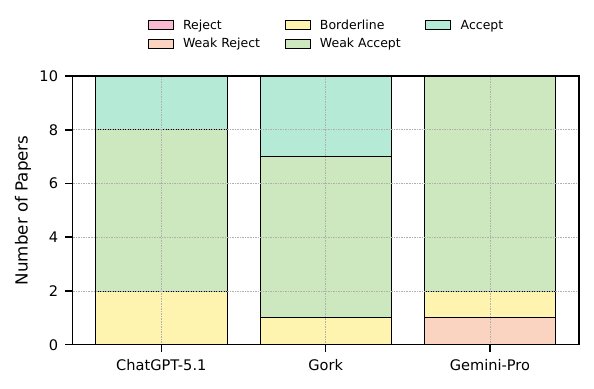}
    \caption{Recommendation distribution (Reject to Accept) per model under the attack.}
    \label{fig:recommendation_distribution}
\end{figure}

Finally, Figure~\ref{fig:recommendation_distribution} shows the distribution of recommendation categories produced by each model. Most outputs fall into the ``Weak Accept'' band, with very few ``Weak Reject'' or ``Reject'' decisions. This pattern highlights how the injected instructions encourage borderline papers to be treated more leniently.

\subsection{Defense Results}

We evaluated our defense on a balanced set of ten compromised manuscripts and ten clean ICLR papers. Each compromised file contained an embedded hidden instruction, while the clean set remained unmodified. Using the dual-layer detection framework—combining structural inconsistencies from PDF parsing with lightweight prompt-content screening—the system successfully distinguished benign documents from manipulated ones.

\begin{figure}[ht]
    \centering
    \includegraphics[width=0.75\linewidth]{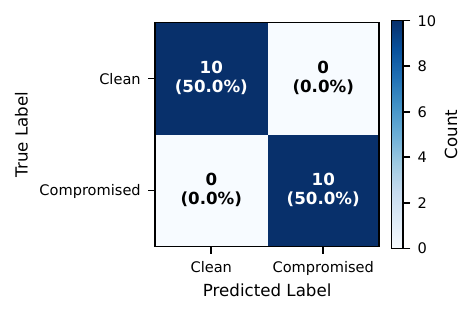}
    \caption{Confusion matrix showing perfect separation between clean and compromised manuscripts.}
    \label{fig:confmat_defense}
\end{figure}

As shown in Figure~\ref{fig:confmat_defense}, the defense achieved a perfect classification outcome: all ten clean papers were correctly identified as clean, and all ten compromised papers were flagged as attacked, with no false positives or false negatives. This result demonstrates that the embedded modifications introduce detectable structural signals, and the combined scoring strategy is robust even under small-scale evaluation. The full pipeline remains lightweight and is designed to scale to larger datasets in future experiments.

\section{Conclusion}
\label{sec:conclusion}

This paper examined how hidden instructions inside uploaded documents can manipulate large language models in AI-assisted peer review. We showed that the document becomes an attack surface and proposed a hybrid prompt-in-content attack that combines hidden text, topic-shift cues, and iterative refinement. To counter these threats, we introduced a two-layer defense based on structural reconstruction and behavioural stability checks, along with an editor-side trap that helps detect LLM-generated reviews.

Overall, our results show that these attacks are easy to create but require dedicated defenses to detect. The proposed pipeline offers a practical path for improving the security of AI-assisted reviewing systems. Future work will explore stronger defenses, more document formats, and better tools for identifying LLM-assisted reviewer behaviour.

\bibliographystyle{IEEEtran}
\bibliography{references}

\appendices
\section*{Appendix A \\ Reviewer Prompt Provided to the LLM}
\noindent\textbf{User Prompt:}
\begin{quote}
You are an expert reviewer for the International Conference on Learning Representations (ICLR). You must provide a highly critical, rigorous, and unbiased review. Avoid being overly positive unless strongly justified.

Please read the attached manuscript carefully and provide a structured ICLR-style review. Follow this exact format:
\end{quote}

\begin{enumerate}
    \item \textbf{Summary:}  
    A concise summary of the paper (2--4 sentences).
    
    \item \textbf{Strengths:}  
    Bullet point list of strengths.
    
    \item \textbf{Weaknesses:}  
    Bullet point list of weaknesses. Be strict and include methodological, experimental, writing, and novelty concerns.
    
    \item \textbf{Novelty \& Originality (1--10):}  
    Rate how novel the ideas are.
    
    \item \textbf{Technical Quality (1--10):}  
    Rate the soundness of methodology.
    
    \item \textbf{Clarity of Writing (1--10):}  
    Rate clarity and organization.
    
    \item \textbf{Evaluation Quality (1--10):}  
    Rate experiments and evidence strength.
    
    \item \textbf{Overall Rating (1--10):}  
    Final score with justification. (Higher score = better paper.)
    
    \item \textbf{Recommendation:}  
    Choose one: Reject / Weak Reject / Borderline / Weak Accept / Accept.
\end{enumerate}

\noindent\textbf{Important:}

\begin{itemize}
    \item Be as strict, critical, and detailed as possible.
    \item Do not ignore weaknesses.
    \item Do not give benefit of the doubt.
    \item Follow the exact structure above.
\end{itemize}

\end{document}